\documentclass[twocolumn,aps,showpacs,superscriptaddress,amsmath,amssymb,preprintnumbers]{revtex4}
\usepackage{amsmath,amssymb,graphicx,afterpage,here}
\usepackage{afterpage,subfigure,epsfig,psfig,rotating,minitoc}
\usepackage[english,activeacute]{babel}
\usepackage{ae}
\usepackage{dcolumn}
\newcommand{\be}{\begin{equation}}
\newcommand{\ee}{\end{equation}}
\newcommand{\bea}{\begin{eqnarray}}
\newcommand{\eea}{\end{eqnarray}}

\newcommand{\pd}{\partial}
\newcommand{\etal}{{\em et al.}}

\newcommand{\C}[1]{\cite{#1}}
\newcommand{\F}[1]{Fig. \ref{#1}}

\newcommand{\LB}[1]{\label{#1}}

\begin{document}


\title{Non-Boussinesq Convection at Low Prandtl Numbers:
Hexagons and Spiral Defect Chaos}

\author{Santiago Madruga\footnote{Present address: 
Max-Planck-Institute for Physics of Complex Systems, D-01187
Dresden, Germany}}
\affiliation{Department of Engineering Sciences and Applied Mathematics,
Northwestern University, Evanston, IL 60208, USA}
\author{Hermann Riecke}
\affiliation{Department of Engineering Sciences and Applied Mathematics,
Northwestern University, Evanston, IL 60208, USA}

\date{\today}



\typeout{ !Sputnik....}

\begin{abstract}

We study the stability and dynamics of non-Boussinesq convection in
pure gases (CO$_2$ and SF$_6$) with Prandtl numbers near $Pr\simeq 1$
and in a H$_2$-Xe mixture with $Pr=0.17$. Focusing on the strongly
nonlinear regime we employ Galerkin stability analyses and direct
numerical simulations of the Navier-Stokes equations. For $Pr \simeq
1$ and intermediate non-Boussinesq effects we find reentrance of
stable hexagons as the Rayleigh number is increased. For stronger
non-Boussinesq effects the hexagons do not exhibit any amplitude
instability to rolls. Seemingly, this result contradicts the
experimentally observed transition from hexagons to rolls. We resolve
this discrepancy by including the effect of the lateral walls.
Non-Boussinesq effects modify the spiral defect chaos observed for
larger Rayleigh numbers. For convection in SF$_6$ we find that
non-Boussinesq effects strongly increase the number of small, compact
convection cells and with it enhance the cellular character of the
patterns. In H$_2$-Xe, closer to threshold, we find instead an
enhanced tendency toward roll-like structures. In both cases the
number of spirals and of target-like components is reduced. We
quantify these effects using recently developed diagnostics of the
geometric properties of the patterns.

\end{abstract}


\maketitle

\section{Introduction}

Rayleigh-B\'enard convection of a horizontal fluid layer heated from
below is a paradigmatic system for the study of complex spatial and
spatio-temporal patterns \C{BoPe00}. Usually, the system parameters
are chosen such that variations of the fluid properties across the
layer are very small. In the theoretical description this allows the
Oberbeck-Boussinesq (OB) approximation, in which the temperature
dependence is neglected in all fluid properties except for the density
term that is responsible for the buoyancy. For large temperature
differences across the layer the variation of the fluid properties
with the temperature become significant. These non-Oberbeck-Boussinesq
(NB) effects  break the up-down symmetry that is characteristic of the
OB approximation and, thus, allow a resonant triad interaction among
three Fourier modes whose wavevectors form a hexagonal planform. Due
to the resonant interaction the primary bifurcation to the hexagons is
transcritical, and hexagons are preferred over rolls in the immediate
vicinity of onset \cite{Bu67}. 

Within the framework of the leading-order weakly nonlinear analysis,
hexagons typically become unstable to rolls further above threshold,
where the amplitudes are larger and the resonant-triad interaction
loses significance compared to interactions involving four modes
\cite{Pa60,SeSt62,Se65,Bu67,PaEl67,DaSe68}. This scenario of a
transition from hexagons to rolls has been confirmed in quite a number
of experimental investigations
\cite{SoDo70,DuBe78,Ri78,WaAh81,BoBr91,PaPe92}, and quite commonly it
has been assumed that hexagon patterns in NB convection are confined
to the regime close to onset.

Two convection experiments using SF$_6$ as the working fluid
\C{AsSt96,RoSt02} have shown, however, that even in the strongly
nonlinear regime stable hexagon patterns can be observed. Under OB
conditions \cite{AsSt96} hexagons were found at relatively high
Rayleigh numbers, $\epsilon \equiv (R-R_c)/R_c \approx 3.5$ 
\C{AsSt96}. Due to the mid-plane symmetry of OB convection hexagons
with up-flow in the center coexist stably with down-flow hexagons  in
this regime. In experiments using SF$_6$ under NB conditions
\cite{RoSt02} the hexagons that appear at threshold were replaced by
rolls for somewhat larger heating and then reappeared in the strongly
non-linear regime near $\epsilon ={\mathcal O}(1)$. The
restabilization was attributed to the large compressibility of SF$_6$
near its critical point \cite{RoSt02}. The hexagons that regain
stability were termed {\em reentrant hexagons}.

Recent numerical computations \cite{MaRi05} have demonstrated that 
hexagons can restabilize in NB convection even if the fluid is {\em
incompressible}.  For instance, in water hexagons that are unstable at
$\epsilon=(R-R_c)/R_c =0.15$ can already restabilize at
$\epsilon=0.2$. The origin of this reentrance was traced back to the
existence of stable hexagons in OB convection at large Rayleigh
numbers, and the dependence of the NB effects on the Rayleigh number.

At low Prandtl numbers ($Pr\simeq 1$) and further above onset OB
convection exhibits a new state: {\em spiral defect chaos} (SDC). It
was first found experimentally  \C{MoBo93,MoBo96,LiAh96} and then
investigated theoretically using simple models \cite{XiGu93,CrTu95} as
well as simulations of the full fluid equations \cite{DePe94a,ChPa03}.
This fascinating state of spatio-temporal chaos is characterized by
rotating spirals with varying numbers of arms and of different size,
which appear and disappear irregularly and interact with each other
and with other defects. SDC arises from the roll state at a threshold 
that can be as low as $\epsilon=0.1$ in the limit of small $Pr$.  It
is driven by large-scale flows that are induced by roll curvature and
have a strength that is proportional to $Pr^{-1}$ \cite{ChPa03}.

So far, strongly non-linear NB convection has been studied mostly for
large Prandtl numbers \cite{YoRi03b,MaPe04,MaRi05}, but little is
known for small ( $Pr\simeq 1$) or very small Prandtl numbers ($Pr\ll 
1$). In particular, whether reentrant hexagons exist at large
$\epsilon$  in the presence of the large-scale flows that arise at low
$Pr$, and how NB effects impact spiral defect chaos are interesting
questions, which we address in this paper. 

Here we study NB convection in gases with small Prandtl numbers.
Specifically, we consider parameters corresponding to convection in
CO$_2$ and SF$_6$ ($Pr\simeq 0.8$) and in a H$_2$-Xe mixture
($Pr=0.17$). We show that reentrant hexagons are possible in
convection in CO$_2$. For spiral defect chaos in SF$_6$ we find that
NB effects promote small convection cells (`bubbles'). In H$_2$-Xe,
closer to threshold, roll-like structures dominate. In both cases the
NB effects reduce the spiral character of the pattern. We quantify the
impact of the NB effects on spiral defect chaos using geometric
diagnostics that we have proposed recently \cite{RiMa06}.  

The paper is organized as follows. In Sec.\ref{sec:basicequations} we
briefly review the basic equations, emphasizing how our computations
focus on weakly non-Boussinesq effects, but strongly nonlinear
convection. In Sec.\ref{sec:co2stability} we present the results for
the linear stability of hexagons and rolls in CO$_2$ for a range of
parameters accessible experimentally. To compare with experiments, we
study the influence of different lateral boundary conditions on the
transition from hexagons to rolls in Sec. \ref{sec:sim-co2}.  In Sec.
\ref{sec:sim-sf6} we discuss  spiral defect chaos in SF$_6$ under NB
conditions.  The stability of hexagons and spiral defect chaos in
fluids with very low Prandlt number ($Pr=0.17$) is studied in a 
mixture of $H_2$ and $Xe$ in Sec. \ref{sec:h2xe}. Finally, conclusions
are drawn in Sec.\ref{sec:conclusions}.

\section{Basic equations \LB{sec:basicequations} }

The basic equations that we use for the description of NB
convection have been discussed in detail previously
\cite{YoRi03b,MaRi05}. We therefore give here only a brief summary. 
We consider a horizontal fluid layer of thickness $d$, density $\rho$,
kinematic viscosity $\nu$, heat conductivity $\lambda$, thermal
diffusivity $\kappa$, and specific heat $c_p$. The system is heated
from below (at  temperature $T_1$) and cooled from above (at
temperature $T_2 < T_1$). 

To render the governing equations and boundary conditions
dimensionless we choose the length $d$, the time $d^{2}/\kappa_0$, the
velocity $\kappa_0/d$, the pressure $\rho_0\nu_0 \kappa_0/d^{2}$, and
the temperature $T_s=\nu_0 \kappa_0/\alpha_0 g d^3$ as the respective
scales. The subscripted quantities refer to the respective values at
the middle of the fluid layer in the conductive state. The
non-dimensionalization gives rise to two dimensionless quantities: the
Prandtl number $Pr=\nu_0/\kappa_0$, and  the Rayleigh number
$R=\alpha_0 \Delta T g d^3/\nu_0 \kappa_0$.  Furthermore, we write the
equations in terms of the dimensionless momentum density $v_i=\rho d
u_i/\rho_0 \kappa_0$ instead of the velocities $u_i$. The
dimensionless form of the temperature $\hat T =T/T_s$, heat
conductivity $\hat \lambda =\lambda/\lambda_0$, density  $\hat \rho
=\rho/\rho_0$, kinematic viscosity $\hat \nu =\nu/\nu_0$, and specific
heat $\hat c_p =c_p/c_{p0}$ will be used in the  ensuing equations and
the hats dropped for clarity.  In dimensionless form the equations for
the momentum, mass conservation and heat are then given, respectively,
by
\bea
\frac{1}{Pr}\left(\pd_tv_i+v_j\pd_j\left(\frac{v_i}{\rho}\right)\right)&=&-\pd_i
p  \\
&&+\delta_{i3}\left(1+\gamma_1(-2 z+\frac{\Theta}{R})\right)\Theta\nonumber \\
&&+\pd_j\left[\nu\rho\left(\pd_i(\frac{v_j}{\rho})+\pd_j(\frac{v_i}{\rho})\right)\right]
\nonumber \LB{e:v}\\
\pd_jv_j&=&0, \LB{e:cont}\\
\pd_t\Theta+\frac{v_j}{\rho}\pd_j\Theta
& =&\frac{1}{\rho
c_p}\pd_j(\lambda\pd_j\Theta)-\gamma_3\pd_z\Theta-\nonumber \\
&& R\frac{v_z}{\rho}(1+\gamma_3z).\LB{e:T}
\eea
with the dimensionless boundary conditions 
\bea
\vec{v}(x,y,z,t)=\Theta(x,y,z,t)=0  \mbox{ at } z= \pm \frac{1}{2}.\LB{e:bc}
\eea
Here $\Theta$ is the deviation of the temperature field from the basic
conductive profile. Summation over repeated indices is assumed.

We consider the NB effects to be weak and retain in a
Taylor expansion of all material properties only the leading-order
temperature dependence {\it beyond} the OB approximation. For
the density this implies also a quadratic term with coefficient
$\gamma_1$. It contributes, however, only to the buoyancy term in
(\ref{e:v}); in all other expressions it would constitute only a
quadratic correction to the leading-order NB effect. Thus,
the remaining temperature dependence of the fluid parameters $\rho$,
$\nu$, $\lambda$, and $c_p$ in (\ref{e:v},\ref{e:cont},\ref{e:T}) is
taken to be linear
\bea
\rho(\Theta)&=&1-\gamma_0(-z+\frac{\Theta}{R}),\LB{e:rhoTh}\\
\nu(\Theta)&=& 1+\gamma_2(-z+\frac{\Theta}{R}),\LB{e:nuTh}\\
\lambda(\Theta)&=&1+\gamma_3(-z+\frac{\Theta}{R}),\LB{e:lambdaTh}\\
c_p(\Theta)&=&1+\gamma_4(-z+\frac{\Theta}{R}).\LB{e:cpTh}
\eea

The coefficients $\gamma_i$ give the difference of the respective
fluid properties across the layer. They depend therefore linearly on
the Rayleigh number, 
\bea
\gamma_i(\Delta T)=\gamma_i^{c}\,\left(\frac{R}{R_c} \right) 
=\gamma_i^{c} \, (1+\epsilon) ,
\eea
where $\gamma_i^{c}$ is the value of $\gamma_i$ at the onset of convection
and $\epsilon\equiv (R-R_c(\gamma_i^{c}))/R_c(\gamma_i^{c})$ is the
reduced Rayleigh number. 

In analogy to \cite{Bu67}, we further omit NB terms that
contain cubic nonlinearities in the amplitudes $v_i$ or $\Theta$, as
they arise from the expansion of the advection terms $v_j
\partial_j(v_i/\rho)$ and $(v_j/\rho)\partial_j \Theta$ when the
temperature-dependence of the density is taken into account.  Since we
will be considering Rayleigh numbers up to twice the critical value,
which implies enhanced NB effects, these approximations
may lead to quantitative differences compared to the fully
NB system, even though the temperature-dependence of the
material properties themselves is in most situations well described by
the linear (or quadratic in the case of the density) approximation.  

To quantify the overall strength of the NB-effects we use Busse's
NB parameter $Q$, which is given by
\bea
Q = \sum_{i=0}^{4}\gamma_i^{c} {\cal P}_i,\LB{e:busseq}
\eea
where the quantities ${\cal P}_i$ are linear functions of $Pr^{-1}$.
The NB parameter $Q$ quantifies the breaking of the up-down symmetry,
which renders at most one of the two types of hexagons stable. Gases
have a positive value of $Q$ and exhibit hexagons with down-flow in
the center ($g$-hexagons), whereas  liquids have negative $Q$ and show
hexagons with up-flow ($l$-hexagons).

We focus in this paper on the stability properties of patterns in the
strongly nonlinear regime. They are determined by a Galerkin method
(e.g. \cite{BuCl79a}). We use a Fourier expansion on a hexagonal
lattice in the lateral directions.  The Fourier wave vectors ${\bf q}$
are constructed as linear combinations  of the hexagon basis vectors
${\bf b}_1 =q(1,0)$ and  ${\bf b}_2 =q(1/2, \sqrt{3}/2)$ with  ${\bf
q} = m {\bf b}_1 + n {\bf b}_2$ where the integers $m$ and $n$ are in
the range  $|m {\bf b}_1+n{\bf b}_2 | \le n_q q$.  The largest
wavenumber is then $n_q q$ and the number of Fourier modes retained is
given by $1+6\sum_{j=1}^{n_q}j$. Typically we use $n_q =3 $. The top
and bottom boundary conditions are satisfied by using appropriate
combinations of trigonometric and Chandrasekhar functions in $z$
(\cite{Ch61,Bu67}). In most of the computations we use $n_z=6$ modes
for each field in Eq.(\ref{e:v},\ref{e:cont},\ref{e:T}). The linear
analysis yields the critical Rayleigh number $R_c$ as well as the
critical wavenumber $q_c$. Both depend on the NB-coefficients
$\gamma_i^{c}$ which in turn depend on $R_c$. Thus, in principle one
obtains an implicit equation for the $\gamma_i^{c}$. The shift in the
critical Rayleigh number away from the classical value $R_c=1708$ due
to the NB-effects is, however, quite small (less than 1 percent) and
therefore the resulting change in the $\gamma_i^{c}$ is negligible. In
this paper we therefore choose the $\gamma_i^{c}$ corresponding to
$R_c=1708$.

To investigate the nonlinear hexagon solutions, we start with
the standard weakly nonlinear analysis to determine the
coefficients of the three coupled amplitude equations for the
modes making up the hexagon pattern.  To obtain the fully
nonlinear solutions requires the solution of a set of nonlinear
algebraic equations for the expansion coefficients with respect
to the Galerkin modes. This is achieved with a Newton solver for
which the weakly nonlinear solutions serve as convenient
starting solutions. In the Galerkin code amplitude instabilities
are tested by linear perturbations of the expansion
coefficients. In addition,  modulational instabilities are considered, which involve the introduction of Floquet
multipliers  $\exp (i {\bf s}\cdot (x,y))$ in the Fourier ansatz  for the linear perturbations 
of the Galerkin solutions. 

We also study the temporal evolution of the system. For that we employ
a Fourier spectral code on a rectangular grid $(i\,dq_x,j\,dq_y)$,
$i,j=1...N$, with $dq_y/dq_x=\sqrt{3}/2$ to accomodate perfect
hexagonal patterns. The same vertical modes are used as in the
Galerkin stability code \cite{DePe94a}. To solve for the time
dependence a fully implicit scheme is used for the linear terms,
whereas the nonlinear parts are treated explicitly (second order
Adams-Bashforth method). The time step is typically taken to be
$t_v/500$, where $t_v$ is the vertical diffusion time. We have tested
that the stability regimes obtained from the Galerkin analysis are
consistent with the direct numerical simulations. Both codes employed
in this paper have been kindly provided by W. Pesch
\cite{DePe94a,Pe96}. 

\section{Reentrant hexagons in CO$_2$ \LB{sec:co2stability}}

In this paper we investigate specific scenarios for convection in
gases that should be experimentally realizable. We focus in this
section on CO$_2$ in a conventional range for the layer thickness,
pressure, and temperature. Table \ref{t:gamma-co2} provides the NB
coefficients and the $Q$-value at the onset of convection for a
representative range of the mean temperature $T_0$ in a layer of
thickness $d=0.08\,cm$ \footnote{These values were obtained with a
code kindly provided by G. Ahlers.}. 

\begin{table}
\begin{tabular}{|c|cccccccc|}\hline 
$T_0$ & $\Delta T_c$ & $Pr$ & $\gamma_0^{c}$ &
$\gamma_1^{c}$ & $\gamma_2^{c} $ & $\gamma_3^{c}$ &
$\gamma_4^{c}$ & $Q$ \\ \hline 
20 & 9.43  & 0.87  & 0.0486 & -0.0669 & 0.0779 &0.0236 & -0.0251&  1.199        \\ \hline   
40 & 15.52 & 0.84  & 0.0685 & -0.0883 & 0.1132 &0.0508 & -0.0184& 1.724       \\ \hline   
60 & 23.80 & 0.82  & 0.0931 & -0.1148 & 0.1566 &0.0919 & -0.0074 & 2.430       \\ \hline   
\end{tabular}  
\caption{Values for the critical temperature (in $^\circ C$), the 
Prandtl number $Pr$, NB coefficients
$\gamma_i^{c}$,  and Busse's parameter $Q$ for CO$_2$ at the
onset of convection for three values of the mean temperature (in
$^\circ C$). The values correspond to a layer thickness of 
$d=0.08\,cm$ and a pressure of $P=300$\,psi.  \LB{t:gamma-co2} } 
\end{table}

\subsection{Amplitude instabilities}

In our analysis we first concentrate on spatially periodic solutions
with their wavenumber being fixed at the critical wavenumber and
discuss their domains of existence and their stability with respect to
amplitude instabilities, which do not change the wavenumber.   For
three different cells with thicknesses $d=0.07,\, 0.08, \mbox{ and }
0.09$\, cm, respectively, we consider the range $0 < \epsilon < 1$ at
a pressure of $P=300$\,psi. 

\begin{center}
\begin{figure}
\centering
\includegraphics[width=0.4\textwidth,angle=0]{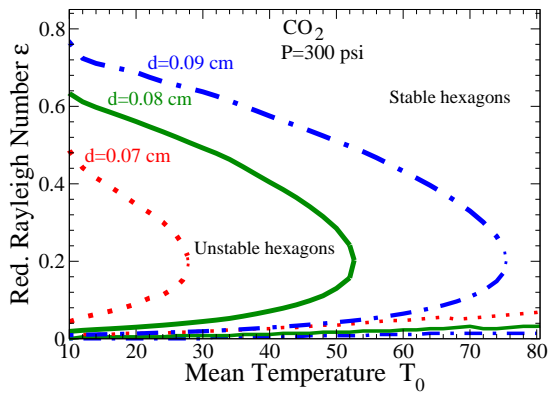}
\caption{Stability regions for hexagons and rolls in CO$_2$ with respect to amplitude 
perturbations for three fluid depths: $d=0.07\, cm$ (dotted
lines), $d=0.08\, cm$ (full lines), $d=0.09\, cm$ (dot-dashed line).
Pressure is kept at $P=300$\,psi.
Thick lines: stability boundaries for hexagons. Thin lines: stability
boundaries for rolls. For a given depth, rolls are stable above the thin 
line, and hexagons are unstable in the inner region of the thick line.  \LB{fig:co2-ampli3d}}
\end{figure}
\end{center}

The results of the stability analysis for hexagons and rolls are shown
in Fig.\ref{fig:co2-ampli3d}. The hexagons are linearly stable for
very small $\epsilon$. For a given layer thickness and not too high
mean temperature $T_0$  the hexagons become unstable as  the control
parameter is increased. The hexagon patterns then undergo  a second
steady bifurcation as the control parameter is increased further and
become stable again. Such restabilized hexagons have been termed {\em
reentrant hexagons} \cite{AsSt96,RoSt02,MaRi05}. As the mean
temperature is increased or the layer thickness is decreased the
critical heating and with it the NB effects increase. This shifts the
point of reentrance to lower $\epsilon$ and the lower stability limit
to higher $\epsilon$, decreasing the $\epsilon$-range over which the
hexagons are unstable, until the two limits merge at a temperature
$T_m$. For $T_0>T_m$ the hexagons are amplitude-stable over the whole
range of $\epsilon$ considered ($0 \le \epsilon \le 1$).

\begin{center}
\begin{figure}
\centering
\includegraphics[width=0.4\textwidth,angle=270]{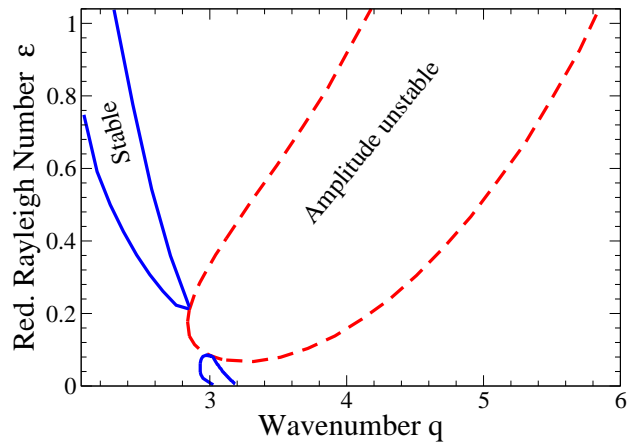}   
\caption{ Stability regions for hexagons in CO$_2$ with respect
to amplitude  (dashed line) and side-band perturbations 
(solid line) for layer depth $d=0.08\,
cm$, pressure $P=300$\,psi, mean temperature $T_0=40\,^oC$, and Prandtl number $Pr=0.84$.
The NB coefficients $\gamma_i^{c}$ are reported in  Tab. \ref{t:gamma-co2}.
Hexagons are stable with respect to amplitude perturbations outside the
dashed-line region, and stable with respect to side-band
perturbation inside the solid-line region.\LB{fig:side-co2}}
\end{figure}
\end{center}

We have also computed the stability of rolls with respect to amplitude
perturbations. The corresponding stability limits are indicated  in
Fig.\ref{fig:co2-ampli3d} by thin lines. Rolls are stable above these
lines. As the NB effects become stronger the stabilization
of rolls is shifted to larger $\epsilon$. In contrast to the
hexagons,  the rolls do not undergo a second bifurcation within the
parameter regime investigated and remain amplitude-stable beyond
$\epsilon =1.0$.  For strong NB
effects one has therefore a large range of parameters over which the
competing rolls and hexagons are both linearly amplitude-stable. 

The amplitude-stability limits of the hexagons and rolls depend on
their wavenumber.  This is illustrated for the hexagons in Fig.
\ref{fig:side-co2} for a mean temperature of $T_0=40\,^o C$. The
instability region with respect to amplitude perturbations forms a
bubble-like closed curve, inside of which the hexagons are unstable
with respect to amplitude perturbations. 

It is worth mentioning that the stability limits for hexagons in
CO$_2$ are quite similar to those of NB convection in
water, except that in CO$_2$ the NB effects increase
rather than decrease with increasing mean temperature \cite{MaRi05}.

\begin{center}
\begin{figure}
\centering
\includegraphics[width=0.4\textwidth,angle=270]{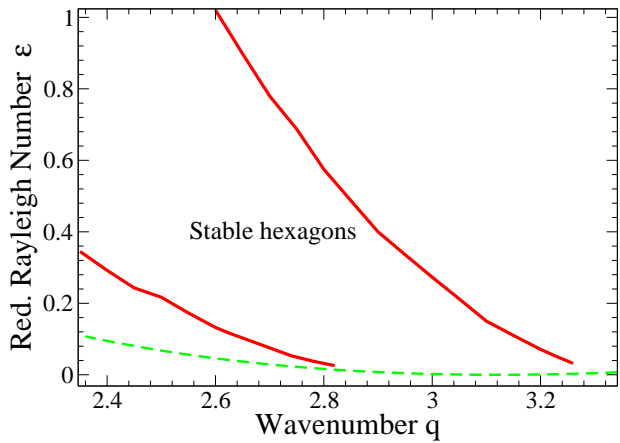}
\caption{ Stability regions for hexagons in CO$_2$ with respect to 
side-band perturbations for layer thickness $d=0.052\, cm$, pressure
$P=300$\,psi, mean temperature $T_0=27^o C$, and Prandtl $Pr=0.86$. The NB coefficients
are $\gamma_0^{c}=0.2053$,  $\gamma_1^{c}= -0.2877$,  $\gamma_2^{c}=0.3267$, $\gamma_3^{c}=0.1152$, and $\gamma_4^{c}=-0.086$.
The hexagons are stable with respect to side-band perturbations in the region 
inside the solid lines and unstable outside. The dashed line corresponds to
the neutral curve. \LB{fig:side-to27}} 
\end{figure}
\end{center}

\subsection{Side-Band Instabilities}

In systems with sufficently large aspect ratio side-band instabilities
can be the most relevant instabilities. Using the Galerkin method, we
have studied the stability of the hexagons with respect to long- and
short-wave side-band perturbations for $T_0=40^oC$. The results are
shown in Fig.\ref{fig:side-co2}. We find that over the whole range
$0\le \epsilon \le 1$ the only relevant side-band perturbations are
long-wave and steady, as is the case in the weakly nonlinear regime.
The same is true also in water with higher Prandtl number \C{MaRi05}.
In this parameter regime the stability region consists of two
disconnected domains, reflecting the reentrant nature of the hexagons.
The stability domain near onset is very small and closes up as the
amplitude stability limit is reached. In the reentrant regime the
stable domain opens up again in an analogous fashion  when the
amplitude-stability limit is crossed. Note that the stability
boundaries are leaning toward lower wavenumbers. Thus, stable
reentrant hexagon patterns are expected to have wavenumbers below
$q_c$. This is related to the fact that in the OB case hexagons can be
stable for large $\epsilon$, but they are side-band stable only for
small wavenumbers \cite{ClBu96}.

As the mean temperature is increased the bubble of the amplitude
instability shrinks and eventually disappears, as shown in
\F{fig:side-to27} for a cell with thickness $d=0.052\, cm$ and mean
temperature $T_0 =27\,^o C$. As before, the relevant side-band
instabilities are long-wave all along the stability limits and with
increasing $\epsilon$ the wavenumber range over which the hexagons are
stable shifts towards smaller wavenumbers. For these parameters the
region of side-band-stable hexagons reaches without interruption from
the strongly nonlinear regime all the way down to threshold. For yet
stronger NB effects the range of stable wavenumbers widens. 

\section{Comparison with experiments in CO$_2$ \LB{sec:sim-co2}}

Bodenschatz \etal \cite{BoBr91} carried out a set of experiments on
convection in CO$_2$ in a cylindrical cell  with  aspect ratio
$\Gamma\sim172$, thickness $d=0.052\, cm$, and pressure $P=300$\, psi.
Under these conditions NB effects are relevant. In the experiments a
weakly hysteretic transition from hexagons to rolls was found near
$\epsilon=0.1$. Noting that this transition point was below the
amplitude instability of hexagons to rolls as predicted by weakly
nonlinear theory, the authors interpreted their results in terms of
the heterogeneous nucleation of rolls by the sidewalls. They found
that for small $\epsilon$ the concentric rolls induced by the sidewall
heating remained confined to the inmediate vicinity of the sidewalls;
however, as $\epsilon$ was increased the rolls invaded the hexagons
and filled the whole cell, inducing a transition from hexagons to
rolls. 

A comparison of the experimental findings with the stability results
shown in Fig. \ref{fig:side-to27} shows that indeed the transition
cannot be due to an amplitude instability of the hexagons. In fact, in
this regime the NB effects are so strong that, in contrast to the
predictions of the weakly nonlinear theory, the hexagons do not
undergo an amplitude instability at all. To clarify the influence of
the sidewalls and to assess the significance of the side-band
instabilities for the transition from hexagons to rolls, we perform
direct simulations of the Navier-Stokes equations
(\ref{e:v},\ref{e:cont},\ref{e:T},\ref{e:bc}) for two different sets
of boundary conditions \footnote{In the experimental setup the top
temperature is held constant at $T=12.84 ^oC$, and therefore the mean
temperature changes as $\epsilon$ is increased. In our computations,
however, we keep $T_0$ fixed. Since the transition occurs quite close
to threshold this is a good aproximation to the experimental
procedure.}: 

i) periodic boundary conditions,

ii) concentric rolls as boundary conditions. In our computations this
type of boundary condition is generated by a suitably patterned 
heating in the interior of the fluid.  In the experiments concentric
rolls near the side walls were generated by a side-wall heating
\cite{BoBr91}.

i) according to Fig. \ref{fig:side-to27}, for $\epsilon=0.3$ hexagons
with wavenumbers larger than $q>2.98$, which includes the critical
wavenumber, are unstable with respect to side-band instabilities. To
test whether these side-band instabilities trigger a transition to
rolls we perform numerical simulations with periodic boundary
conditions  and hexagons as initial conditions. Fig.
\ref{fig:perio-co2} presents some snapshots of the ensuing temporal
evoluation in a cell of size $L=8\cdot 2\,\pi/q_c=16.11$ using
$128\times 128$ Fourier modes. More precisely, to allow perfect
hexagons in a rectangular container the container cannot be square. In
our simulations we use $L_x=L$ and $L_y=\sqrt{3}L/2$. The sideband
instability of the initially almost perfect hexagon pattern (cf.
\F{fig:perio-co2}(a)) induces a shearing of the pattern (cf.
\F{fig:perio-co2}(b)). At the same time a few  penta-hepta defects
arise and some hexagonal convection cells tend to connect with each
other forming short roll-like structures. However, as time evolves the
system becomes progressively more ordered again and eventually after
losing a number of convection cells a defect-free hexagon pattern with
a smaller wavenumber is established (cf. \F{fig:perio-co2}(c) at 
$t\simeq 60t_v$).  

Thus, while roll-like features appear for this value of $\epsilon$,
with periodic boundary conditions no transition to rolls occurs and
the system relaxes to a new ordered hexagon pattern. Only for yet
larger values of $\epsilon$ the roll-like structures that arise in the
intermediate, disordered state take over and lead to a transition to
rolls induced by the sideband instabilities.

\begin{center}
\begin{figure}
\centering
\includegraphics[width=0.5\textwidth]{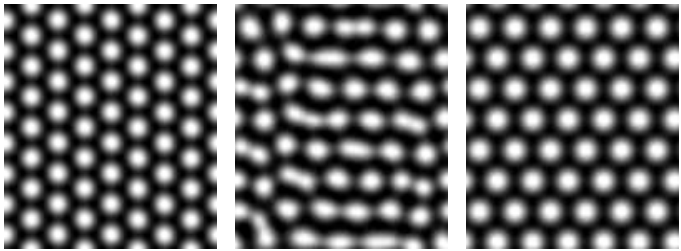}
\caption{ Numerical simulation corresponding to a layer of 
CO$_2$ with thickness $d=0.052\,cm$, pressure $P=316$\,psi, 
mean temperature $T_0=27.3^o C$,  Prandtl number $Pr=0.87$, and control parameter  $\epsilon=0.3$
(cf. Fig.\ref{fig:side-to27}). The size of the integration domain is
$L=8\cdot 2\,\pi/q_c=16.11$ and  the boundary conditions are
periodic.  As initial condition a perfect hexagon lattice with 
wavenumber $q_c=3.12$ has been used.  a) corresponds to $t=0$, 
b) to  $t=18t_v$, and c) to $t=60t_v$.\\ \LB{fig:perio-co2}} 
\end{figure}
\end{center}

ii) clearly, the simulations for periodic boundary conditions do not match
the experimental results described above, where a transition from
hexagons to rolls occurs already for $\epsilon \gtrsim 0.11$. To
address this disagreement we take into account the fact that in the
experiments the side walls promote the nucleation of rolls
\cite{BoBr91}.

Strictly speaking, the code we are using does not allow non-periodic
boundary conditions. To mimic the experimentally used clylindrical
cell we employ a step ramp in $\epsilon$ that reduces $\epsilon$ to
values well below $\epsilon=0$ outside a circle of radius $r=0.45L$
with $L=16\cdot 2\pi/q_c=32.22$ \C{DePe94a}. To induce concentric
rolls near the sidewalls we introduce for $r> 0.45L$ an additional
heating in the interior of the fluid in the form of concentric rings.
Using hexagonal initial conditions in the bulk, this leads to an
initial state as shown in Fig. \ref{fig:boun-co2}a.

Fig. \ref{fig:boun-co2}a,b shows two snapshots at $t=t_v$ and
$t=158t_v$ demonstrating how the rolls induced by the side walls
invade the carefully prepared hexagonal state in the bulk already for
$\epsilon=0.2$. This is well below the $\epsilon$-value for which with
periodic boundary conditions the hexagons persisted even through the
side-band instability. The final steady state consists of concentric
rolls as observed in the experiments  (cf. Fig. 5 in \cite{BoBr91}).
For lower $\epsilon$, however, the experimentally observed final state
consists of hexagons in the bulk of the system surrounded by
wall-induced concentric rolls (cf. Fig. 4 in \cite{BoBr91}). We find
this also in our numerical simulations, as shown in
\F{fig:boun-co2}(c-d). There the forcing of rolls is identical to that
in \F{fig:boun-co2}(a-b) but $\epsilon=0.05$. Starting with random
initial conditions, the forcing gives rise to a ring contained in the
square integration domain.  At the beginnig of the simulation
(\F{fig:boun-co2}(c)) the rolls created by the forcing invade the
interior of this small system. However, as time progresses the rolls
pull out of the central region of the cell,  and the  final steady
state  (for $t \gtrsim  165\,t_v$)  consists of stable hexagons
surrounded by a couple of concentric rolls in addition to those 
induced by the forcing. 

\begin{center}
\begin{figure}
\centering
\includegraphics[width=0.4\textwidth]{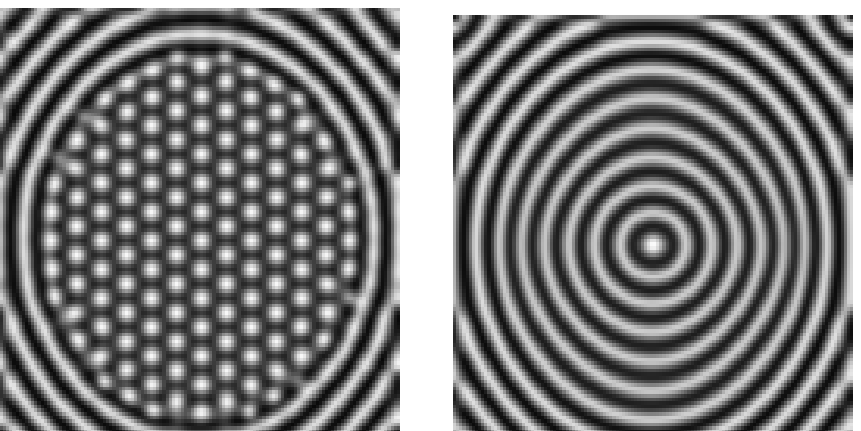}\\
\vspace{0.5cm}
\includegraphics[width=0.4\textwidth]{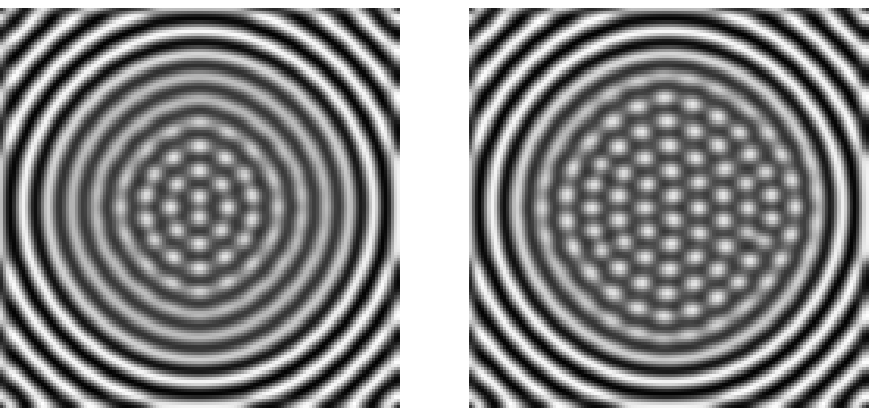}
\caption{ Numerical simulation in a cell of CO$_2$ with thickness
$d=0.052\,cm$, pressure $P=300$\,psi, mean temperature $T_0=27^o
C$, and  Prandtl number $Pr=0.86$. The size of the integration domain
is $L=16\cdot 2\,\pi/q_c=32.22$, and the boundary conditions are rings
of concentric rolls  generated with and external forcing. \\ For
$\epsilon=0.2$ (upper snapshots) hexagon inititial conditions have
been used, with $t=t_v$ (left snapshot) and $t=158\,t_v$ (right). The
lower snapshots correspond to a simulation with $\epsilon=0.05$ and
random initial conditions, at $t=40\,t_v$ 
(left) and $t=158\,t_v$ (right).\LB{fig:boun-co2}} 
\end{figure}
\end{center}

Thus, our simulations suggest that the experimentally observed
transition from hexagons to rolls is neither due to amplitude
instabilities nor to side-band instabilities. Rather, there is a large
range of parameters in which hexagons and rolls are both linearly
stable and the final state is selected by one type of pattern invading
the other. The transition to rolls at these low values of $\epsilon$
is made possible by the boundaries, which provide a seed for the
nucleation of rolls. We expect that by applying a forcing that is
confined to the region near the boundaries and that replaces the rolls
by hexagons the transition to rolls could be shifted to substantially
larger values of $\epsilon$. Such a forcing could be achieved by a
patterned heating of the interior of the fluid \cite{SeSc02} or
possibly by a suitable geometric patterning of the bottom plate
\cite{Bopriv}.

\section{NB Spiral Defect Chaos in SF$_6$\LB{sec:sim-sf6}}

A fascinating state observed in convection at low Prandtl numbers is
spiral defect chaos. It is characterized by self-sustained chaotic
dynamics of rotating spirals, as well as dislocations and
disclinations, and arises in fluids  with $Pr\lesssim 1$
\C{MoBo93,XiGu93,DePe94a,CrTu95,LiAh96} in a parameter range where
straight rolls are linearly stable. Spiral defect chaos has so far
predominantly been investigated under OB conditions, in which
up-flows and down-flows are equivalent. 

As mentioned before, NB effects break the up-down symmetry and
different flow structures may be predominantly associated with up-flow
and down-flow, respectively. Moreover, in the absence of the OB
symmetry a resonant triad interaction is allowed. If it is strong
enough it leads to the formation of hexagons. For weaker interaction
one may still expect an enhancement of cellular rather than roll-like
structures.

To investigate the impact of NB effects on spiral defect chaos we
consider convection in a layer of SF$_6$. This gas has been used
previously in experimental convection studies under usual laboratory
conditions \C{BoCa92}, and near the thermodynamical critical point
\C{AsSt94,AsSt96,RoSt02}. In Fig. \ref{fig:sf6-ampli}(a) we present
the stability diagram for hexagons and rolls with respect to amplitude
perturbations in a layer of SF$_6$ of  thickness $d=0.0542$\,cm,
pressure $P=140\,$psi, and a range of temperatures that is
experimentally accessible. Hexagons are amplitude-stable to the right
of the solid line and rolls above the dashed line.  As in the case of
CO$_2$ the NB effects increase with increasing mean temperature $T_0$
and above a certain value of $T_0$ hexagons are linearly
amplitude-stable over the whole range of $\epsilon$ investigated. Here
we focus on relatively strong NB effects. We therefore show in
Fig.\ref{fig:sf6-ampli}b the stability limits with respect to
side-band perturbations for a relatively large mean temperature,
$T_0=80^\circ C$. As in the case of CO$_2$, the wavenumber range over
which the hexagons are stable is leaning towards smaller wavenumbers.
Overall, amplitude and side-band stability limits are qualitatively
similar to those of convection in CO$_2$ (cf. Fig.
\ref{fig:co2-ampli3d}).

\begin{center}
\begin{figure}
\begin{minipage}{0.35\textwidth}
\includegraphics[width=\textwidth,angle=0]{sf6.eps}
\end{minipage} 
\hspace{0.4cm}  
\begin{minipage}{0.35\textwidth}
\includegraphics[width=\textwidth,angle=270]{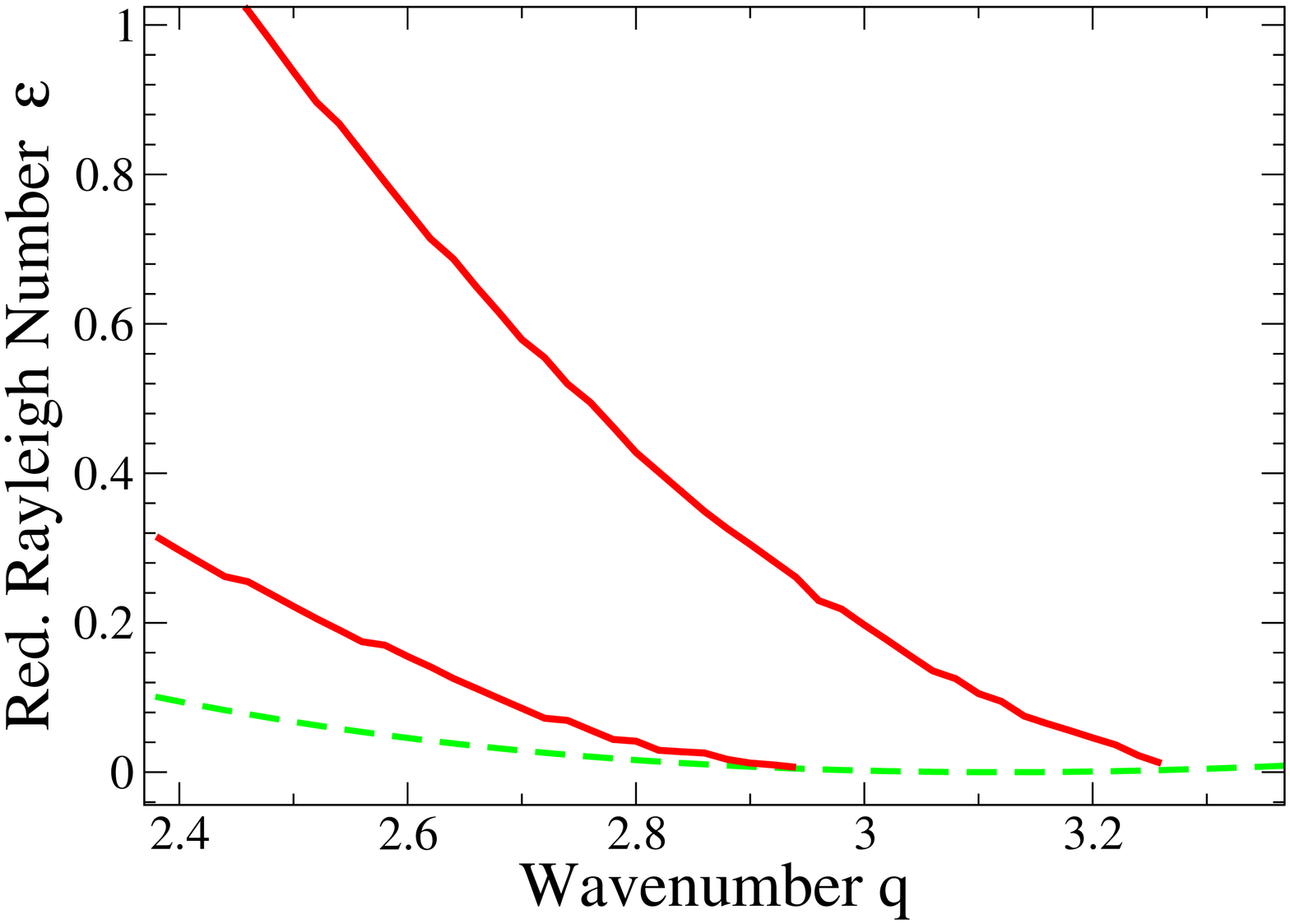}
\end{minipage}
\caption{ 
Stability regions of hexagons and rolls in a layer of  SF$_6$ 
of thickness $d=0.0542\,cm$,  pressure $P=140$\,psi, and Prandtl number $Pr=0.8$. \\
a)  Stability regions with respect to amplitude perturbations. 
Continous line: stability boundary for hexagons. Dashed line: stability boundary for
rolls. Stability limits obtained for the critical wavenumber $q_c$.\\
b)  Stability regions with respect to side-band perturbations for the above
layer with a mean temperature $T_0=80\,^oC$ The corresponding NB coefficients
are $\gamma_0^{c}=0.1714$,  $\gamma_1^{c}= -0.2118$, $\gamma_2^{c}=0.2836$, $\gamma_3^{c}=0.1905 $, and
$\gamma_4^{c}=0.0624$ corresponding to $Q=4.2$.. The dashed line corresponds to
the neutral curve. 
\LB{fig:sf6-ampli}}
\end{figure}
\end{center}

Fig.\ref{fig:sf6-nb-ob} shows two snapshots obtained by direct 
numerical simulations of the Navier-Stokes equations corresponding to
convection in SF$_6$ for $T_0=80\,^oC$ and $\epsilon=1.4$ in a
convective cell of thickness $d=0.0542\,cm$ and horizontal size
$L=16\cdot 2 \,\pi/q_c=32.22$. Periodic boundary conditions are used
with $128\times 128$ Fourier modes and 6 vertical modes.  Both states
are obtained after an integration time of $160\,t_v$, starting from
random initial conditions. While in
Fig.\ref{fig:sf6-nb-ob}b all NB are retained, in
Fig.\ref{fig:sf6-nb-ob}a the same values are used for $Pr$ and
$\epsilon$, but all NB parameters $\gamma_i^c$ are set to 0, i.e. the
system is treated as if it was Boussinesq. 

\begin{center}
\begin{figure}
\centering
\includegraphics[width=0.4\textwidth]{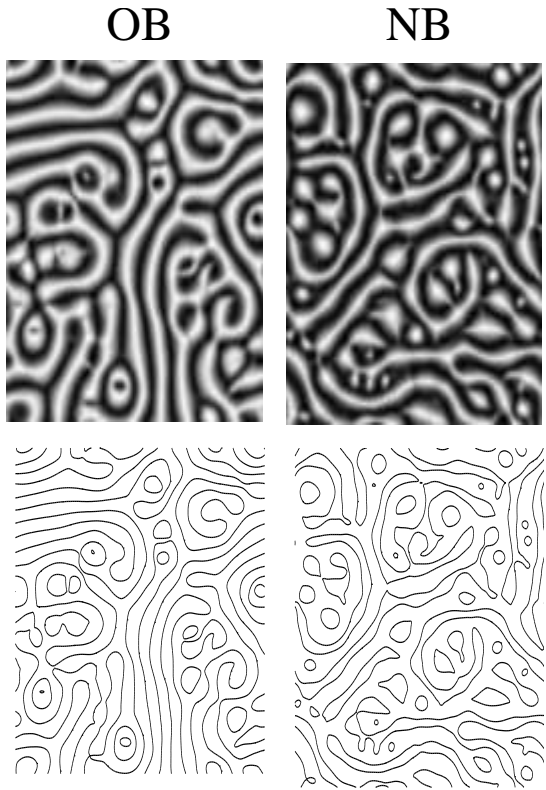}
\caption{ 
Direct numerical simulation of equations (\ref{e:v}-\ref{e:bc}) for SF$_6$ in a
cell of thickness $d=0.0542\,cm$, pressure $P=140$\,psi, mean temperature 
$T_0=80^oC$, Prandtl number $Pr=0.8$, and control parameter $\epsilon=1.4$. 
The cell has size $L=16\cdot 2 \,\pi/q_c=32.22$ with periodic boundary conditions. 
Starting from random initial conditions both snapshots are taken at
$t=160\,t_v$. Left panels for OB conditions ($\gamma_i$=0, $i=0..4$, left), 
right panels for NB conditions appropriate for $T_0=80^oC$ 
(cf. Fig\ref{fig:sf6-ampli}b). Bottom panels give the corresponding contour lines used
for the pattern diagnostics.
\LB{fig:sf6-nb-ob}}
\end{figure}
\end{center}

The snapshots depicted in Fig.\ref{fig:sf6-nb-ob} show, as expected,
that due to the NB effects down-flow convection cells, which are white
in Fig.\ref{fig:sf6-nb-ob}, outnumber cells with up-flow (black).
Moreover, in this regime the NB effects enhance the overall cellular
rather than roll-like character of SDC.  This manifests itself in the
appearance of numerous small down-flow convection cells (white
`bubbles') and in the appearance of quite noticeable bulges on the NB
convection rolls. To quantify these and other differences  we analyse
a long sequence of snapshots with a recently introduced geometric
approach \cite{RiMa06}. It is based on the contour lines corresponding
to the intensity half-way between the minimal and maximal intensity of
all snapshots in a given run.  The contour lines corresponding to the
temperature field of snapshots Fig.\ref{fig:sf6-nb-ob}(a,b) are shown in
Fig.\ref{fig:sf6-nb-ob}(c,d). In the following we
present various statistics of these contour lines.

\begin{center}
\begin{figure}
\centering
\includegraphics[width=0.4\textwidth,angle=0]{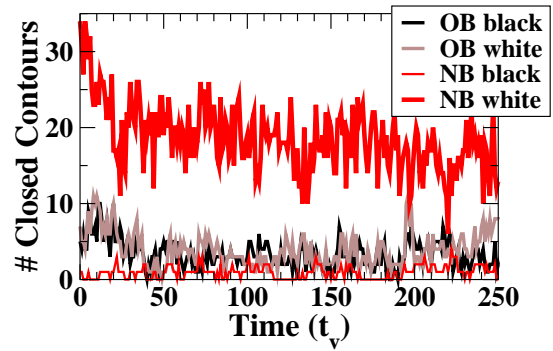} \\
\caption{ 
Number of black and white closed contours  as a function of time
for NB and OB conditions for the simulations corresponding to the
snapshots shown in \F{fig:sf6-nb-ob}. 
\LB{fig:num-bubbles}}
\end{figure}
\end{center}

The most striking difference between the OB and the NB case is the
asymmetry that is induced by the NB effects between `black' and
`white' components, i.e. between closed contours that enclose up- and
down-flow regiones, respectively. To quantify this asymmetry we
measure the number of white and black components. These two
topological measures correspond, respectively, to the Betti numbers of
order 1 and 2 of the pattern defined by the white components
\cite{GaMi04}. Fig.\ref{fig:num-bubbles} shows these two quantities as
a function of time in the OB and the NB case. As expected, in the OB
case the number of black and white components is essentially the same
at all times, whereas in the NB case the white components
significantly outnumber the black ones. The ratio of white to black
components is therefore a sensitive indicator for the significance of
NB effects. Fig.\ref{fig:num-bubbles} also illustrates how much the
number of components fluctuates during these runs. Recently, the two
Betti numbers have also been measured based on patterns obtained in
experiments on SDC in convection in CO$_2$. Scanning $\epsilon$ in
very small steps the authors report steps in the Betti numbers
indicative of transitions between different chaotic, disordered states
\cite{KrGaunpub}.

Fig.\ref{fig:num-bubbles} shows that the total number of components
(closed contours) is considerably larger in the NB case than in the OB
case. This is presented in more quantitative detail in
Fig.\ref{fig:num-bubbles-eps}, which gives the mean value of the total
number of components, i.e. the sum of black and white components, as a
function of $\epsilon$ for OB as well as NB conditions. In the NB case
the total number of components is up to four times larger than in the
OB case. We attribute this difference to the resonant triad
interaction that is made possible by the breaking of the OB symmetry
and which tends to enhance cellular rather than filamentary roll-like
structures.

\begin{center}
\begin{figure}
\centering
\includegraphics[width=0.4\textwidth,angle=0]{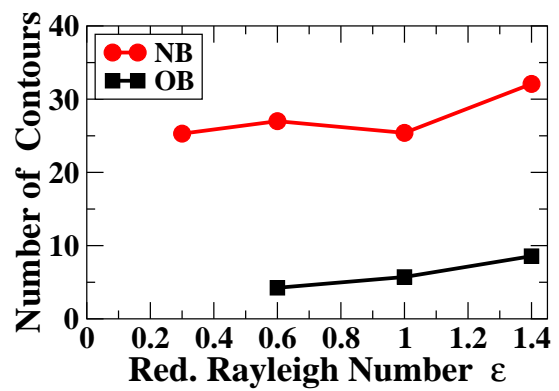} 
\caption{ 
Total number of closed contours as a function of the control parameter 
$\epsilon$ for NB (circles) and OB (squares) conditions. 
\LB{fig:num-bubbles-eps}}
\end{figure}
\end{center}
 
To characterize the components better and to distinguish cellular and
roll-like structures we introduce the `compactness' ${\mathcal C}$ of
components \cite{RiMa06},
\bea
{\mathcal C}=4\pi \frac{{\mathcal A}}{{\mathcal P}^2}.  \LB{e:compact}
\eea 
Here ${\mathcal A}$ is the area inside a closed contour and $P$ its
perimeter. With the normalization used in (\ref{e:compact}) compact,
cellular structures are charecterized by ${\mathcal C}\lesssim 1$,
whereas filamentary, roll-like structures have ${\mathcal C} \ll 1$.

\begin{center}
\begin{figure}
\centering
\includegraphics[width=0.4\textwidth,angle=0]{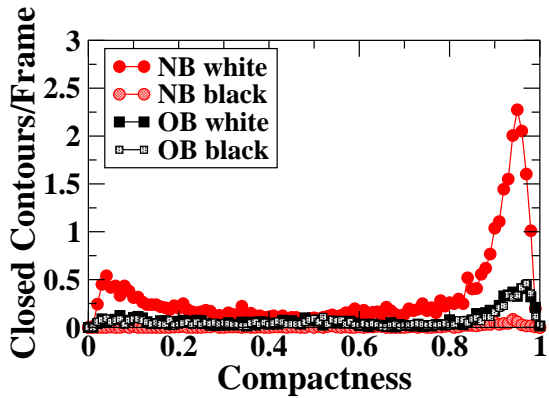}
\caption{ 
Mean number of closed contours per snapshot with a 
given compactness ${\mathcal C}$ for convection in  SF$_6$ at
$\epsilon=1.4$ under NB (circles) and under OB (squares) conditions 
(cf. \F{fig:sf6-nb-ob}).
\LB{fig:hist-compact}}
\end{figure}
\end{center}

\F{fig:hist-compact} shows the mean number of closed contours per
snapshot for a given compactness $\mathcal C$ for the NB and the OB
simulation at $\epsilon=1.4$ over the duration $t=360\,t_v$. As
expected, in the NB case the number of white components is much larger
than that of black components, whereas in the OB case both are about
the same. The total number of components is noticeably larger in the
NB case, which also shows  an increase in the number of white,
filamentary contours with small compactness. 

\begin{center}
\begin{figure}
\centering
\includegraphics[width=0.4\textwidth,angle=0]{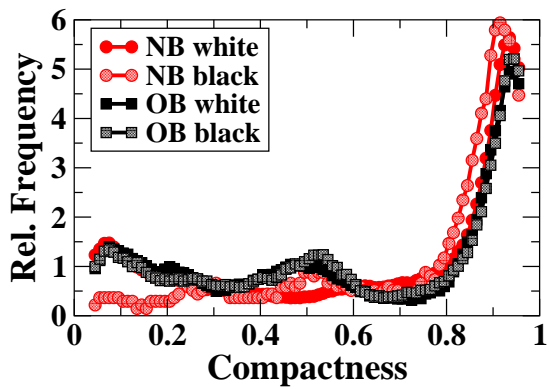}
\caption{ 
Distribution function for the compactness ${\mathcal C}$ for NB
(circles) and OB (squares) conditions (cf. \F{fig:sf6-nb-ob}).
\LB{fig:hist-compact-normalized}}
\end{figure}
\end{center}

The relative distribution among components with different compactness
is more clearly visible in the {\it relative} frequency of contours as
a function of the compactness, which is shown in
Fig.\ref{fig:hist-compact-normalized}. More precisely, its data result
from running averages over adjacent 10 points of the corresponding
data shown in Fig.\ref{fig:hist-compact}, which are then normalized.
The normalized data show that the increase in the relative frequency
of white filamentary components (${\mathcal C}\ll 1$) is essentially
the same in the NB case and in the OB case. However, only very few
black filamentary components arise in the NB case. 

A feature of Fig.\ref{fig:hist-compact-normalized} that is surprising
at first sight is the essentially equal height of the NB peak and the
OB peak for components with ${\mathcal C}\sim 1$. Visually, the NB
snapshot exhibits many more small compact `bubbles' than the OB run.
An explanation for this observation can be obtained by correlating the
compactness of the closed contours with their length ${\mathcal P}$.
The joint distribution function for these two quantities is shown in
Fig.\ref{fig:corr-cont-comp} using logarithmic scales for ${\mathcal
P}$ and ${\mathcal C}$. Focussing on the compact objects with
${\mathcal C}\lesssim 1$ one recognizes in the OB case a second peak
at somewhat larger contourlength. We associate this peak with the
appearance of target-like structures, i.e. with a second contourline
encircling a smaller compact, almost circular contourline. In the NB
case this second peak is barely visible. Instead, the shoulder of the
main peak is extended significantly towards smaller contourlength
${\mathcal P}$. It signifies the appearance of compact objects that
are smaller than a typical wavelength, which we associate with the
small `bubbles' that are easily recognized in the snapshot of the NB
case Fig. \ref{fig:sf6-nb-ob}b. Thus, the comparable relative
frequency of small components shown in
Fig.\ref{fig:hist-compact-normalized} has a different origin in the OB
and the NB case. Whereas in the NB case it is mostly due to small
bubbles, it seems to originate from target-like structures in the OB
case. 

In the logarithmic scaling used in Fig.\ref{fig:corr-cont-comp} a
straight ridge arises in the distribution function for large
contourlengths. It is characteristic for filamentary structures with a
typical width, which corresponds here to half a wavelength $\lambda$
of the convection rolls,
\bea
{\mathcal C}\sim 4\pi
\frac{\lambda \, {\mathcal P}}{4{\mathcal P}^2} \propto {\mathcal
P}^{-1}.
\eea
In the OB case one can discern deviations from this scaling. They are
confined to larger rather than smaller compactness values for a given
contourlength, indicating that the long rolls can be wider but not
narrower than a certain thickness. In the NB case these deviations are
much smaller. 

\begin{figure}
\includegraphics[width=0.48\textwidth]{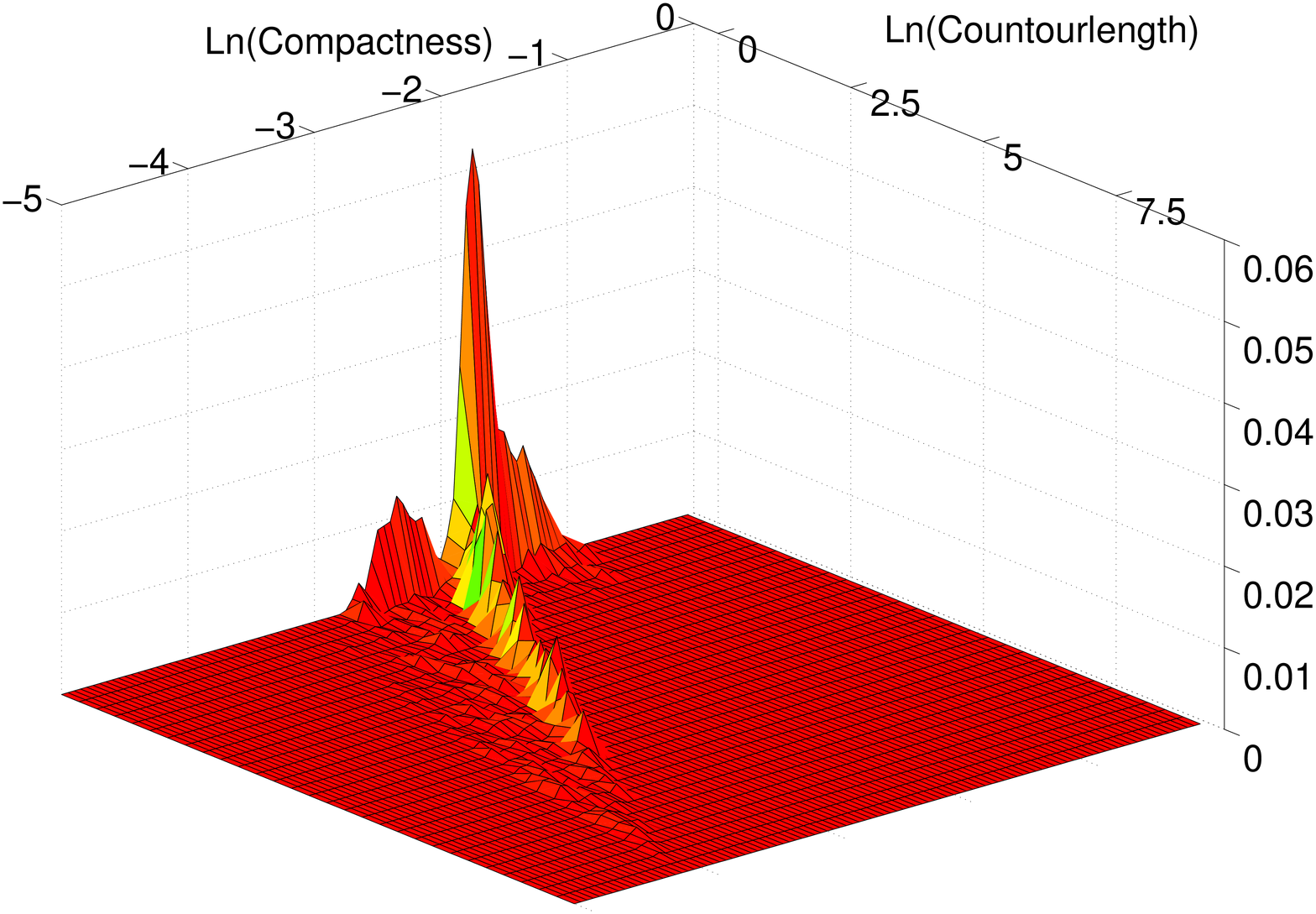}
\includegraphics[width=0.48\textwidth]{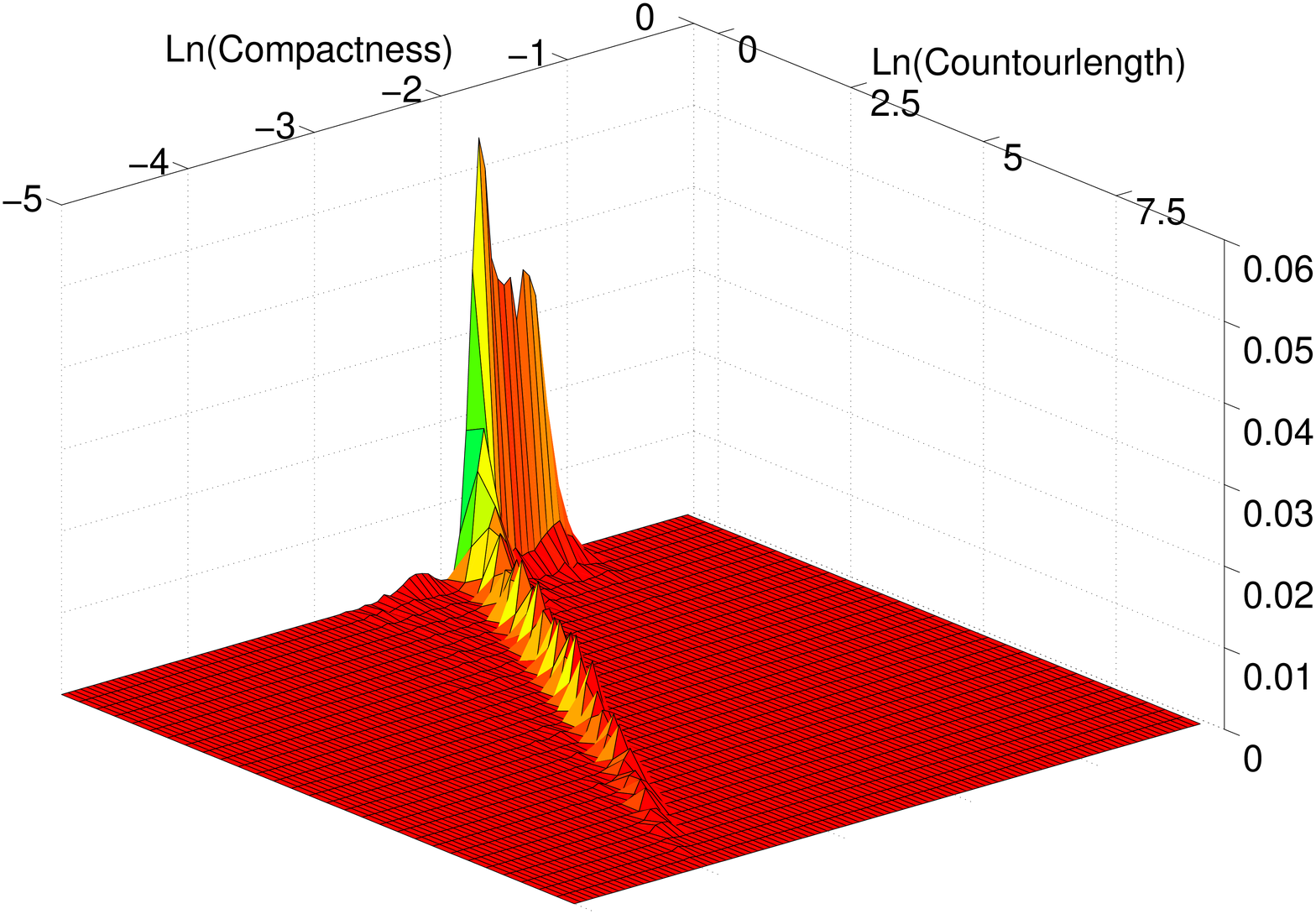}
\caption{Distribution function for contour length and 
compactness for OB (a) and NB conditions (b)
$Pr=0.8$ and $\epsilon=1.4$.
\LB{fig:corr-cont-comp}
} 
\end{figure}

To identify spiral components in the pattern directly we also measure
the winding number of the components \cite{RiMa06}. It is defined via
the angle $\theta$ by which the (spiral) arm of a pattern component is
rotated from its tip to its end at the vertex at which it merges with
the rest of the component. In cases in which a component has no
vertices we split it into two arms at the location of minimal
curvature \cite{RiMa06}. The winding number is then defined as
$|{\mathcal W}|=\theta/2\pi$. To assess the impact of the NB effects
on the spiral character of the pattern we measure the number of
spirals in each snapshot and show the resulting histogram over the
whole run in Fig.\ref{fig:num-spiral}. We use three different
thresholds ${\mathcal W}_{min}$ for the identification of spirals,
${\mathcal W}_{min}= 1$, ${\mathcal W}_{min}= 1/2$, and ${\mathcal
W}_{min}= 1/4$. As  Fig.\ref{fig:num-spiral} shows, the number of
small spirals with $|{\mathcal W}|\gtrsim 1/4$ is quite similar in the
OB and the NB case. However, larger spirals with  $|{\mathcal W}|\ge
1/2$ or even $|{\mathcal W}|\ge 1$ are much more rare in the NB case;
in fact, for the system size $L=16 \cdot 2\pi/q_c=32.22$ that we have
used in these simulations there was at most one spiral with 
$|{\mathcal W}| \ge 1$ at any given time. 

\begin{figure}
\includegraphics[width=0.48\textwidth]{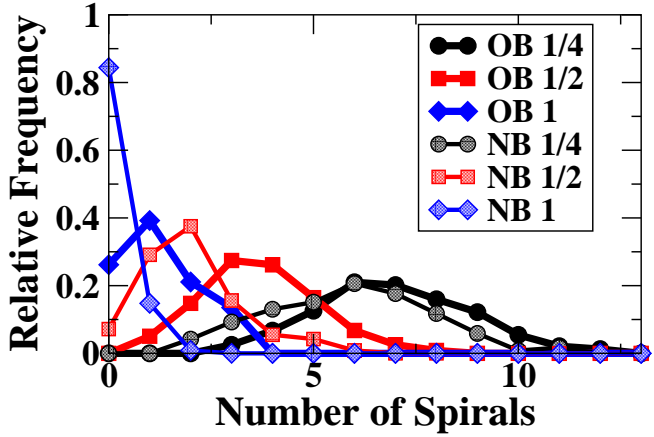}
\caption{Distribution function for the number of spirals in spiral
defect chaos under OB and NB conditions for three values of the
threshold, ${\mathcal W}_{min}=1/4$ (circles),${\mathcal W}_{min}=1/2$
(squares), and ${\mathcal W}_{min}=1$ (diamonds).  
\LB{fig:num-spiral}
} 
\end{figure}

The reduced spiral character of NB spiral defect chaos is also quite
apparent in Fig.\ref{fig:corr-arc-winding}, which shows the
correlation between the winding number and the arclength of the spiral
arm. More precisely, each dot marks the occurrence of one spiral arm
in a snapshot. In the OB case one can see quite clearly a maximal
winding number for any given arclength, which is consistent with an
Archimedean shape of the spiral \cite{RiMa06}\footnote{Note that
detailed analyses of large spirals show deviations from the
Archimedean shape due to the dislocations that accompany finite
spirals \cite{Pl97}}. In the NB case, however, only components with
very small contourlength reach the Archimedean limit and most
components have winding numbers that are much smaller, i.e. the
components are quite straight.

\begin{figure}
\includegraphics[width=0.48\textwidth]{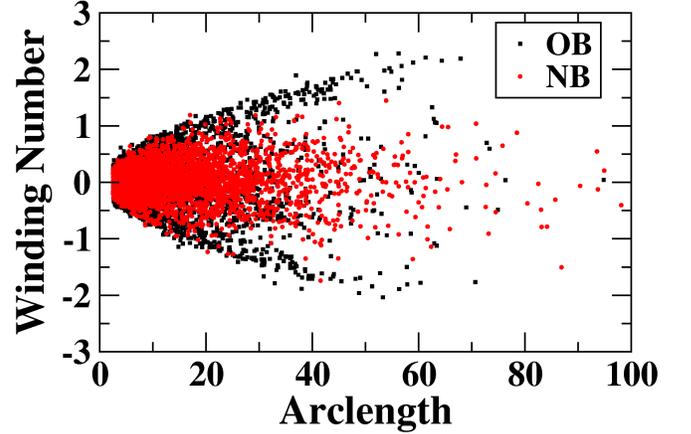}
\caption{Correlation between the arclength and the winding number of
spiral arms under OB (black squares) and NB conditions (red circles).
\LB{fig:corr-arc-winding}} 
\end{figure}

\section{Hexagons and Spiral Defect Chaos at Very Low Prandtl numbers. H$_2$-Xe
Mixtures \LB{sec:h2xe}}

As mentioned above, the restabilization of NB hexagons at larger
Rayleigh number is related to the existence of stable OB hexagons at
large Rayleigh numbers. The wavenumber range over which the OB
hexagons are stable shrinks with decreasing Prandtl number and for $Pr
< 1.2$ the OB hexagons are side-band unstable at {\it all} wavenumbers
\cite{ClBu96}. However, as seen in the case of CO$_2$ and SF$_6$, NB
hexagons can be side-band stable at large $\epsilon$ even below
$Pr=1.2$ due to the additional stabilizing effect of the resonant
triad interaction. It is therefore of interest to investigate whether
the NB effects can be sufficient to stabilize strongly nonlinear
hexagons even for Prandtl numbers significantly below $Pr=1$. 

Prandtl numbers well below $Pr=1$ can be reached by using a mixture of
a heavy and a light gas. An experimentally investigated case is a
mixture of H$_2$ and Xe \cite{LiAh97}. With a mole fraction of
$\chi=0.4$, one can reach Prandtl numbers as small as $Pr=0.17$. The
Lewis number of such a mixture is close to one \cite{LiAh96,Ah05}. 
Therefore such mixtures are expected to behave essentially like a pure
fluid with the same Prandtl number \cite{BoPe00}.

We investigate the stability of hexagon and roll convection in a
H$_2$-Xe mixture with mole fraction $\chi=0.4$ at a pressure of $300$
\,psi and a layer thickness of $d=0.1$cm. With respect to amplitude
instabilities the stability diagram is very similar to that of
convection in CO$_2$ and SF$_6$ with hexagons becoming reentrant at
$\epsilon$-values as low as $\epsilon=0.14$ for $T_0=20\,^oC$. 

Focusing on strong NB effects we perform a detailed stability analysis
with respect to sideband perturbations at a mean temperature of
$T_0=80^oC$ using the same layer thickness of $d=0.1$cm.  For these
low Prandtl numbers the numerical resolution has to be increased to
obtain sufficiently well resolved flows. While for the stability
anlyses of hexagons in CO$_2$ and SF$_6$ it is sufficient to use  
$n_z=6$ and $n_q=3$ in the Galerkin expansion, for the H$_2$-Xe
mixture with $P_r\simeq0.17$ at least $n_q=5$ and $n_z=6$ are
required. \F{fig:side-h2xe}  depicts the resulting stability diagram. It shows
that the region of side-band stable hexagons is not contiguous but
consists of the usual region immediately above threshold and an
additional, disconnected region at larger Rayleigh numbers. We could
not follow the stability limits to smaller values of $q$ than shown in
Fig.\ref{fig:side-h2xe} due to numerical convergence problems.
Presumably, these arise due to bifurcations involving additional,
resonant wavevectors \cite{Mo04a}, somewhat similar to the resonances
studied in Taylor vortex flow \cite{RiPa86,PaRi90}. The fact that the
region of stability is disconnected is remarkable since the region of
amplitude-stability (not shown) is actually contiguous. This is in
contrast to the behavior found in CO$_2$ and SF$_6$ where the two
side-band stable regions become connected when the bubble-like region
of amplitude instability disappears (cf.
Fig.\ref{fig:side-co2},\ref{fig:side-to27}). The comparison of the
stability limits with those of CO$_2$ and of SF$_6$ 
(Fig.\ref{fig:sf6-ampli}b) shows further that the maximal wavenumber
$q$ at which the hexagons are stable with respect to side-band
perturbations decreases with decreasing Prandtl number and as a result
the over-all stability region shrinks as well. 


\begin{center}
\begin{figure}
\centering
\includegraphics[width=0.4\textwidth,angle=270]{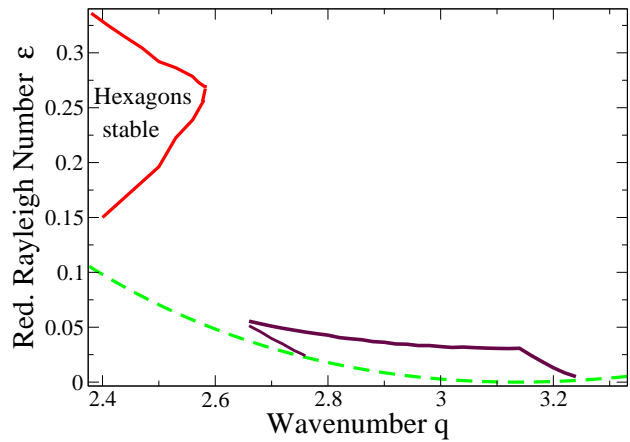}
\caption{ 
Stability limits with respect to sideband perturbations  in  a mixture
of $H_2-Xe$ with molar fraction $\chi=0.4$, thickness $d=0.1 cm$, pressure $P=300$\,psi, 
mean temperature $T_0=80\,^oC$,  and Prandtl number $Pr=0.17$. The NB coefficients are 
$\gamma_0=0.5535$, $\gamma_1=-0.6421$, $\gamma_2=0.9224$, $\gamma_3=0.3647$, 
$\gamma_4=-0.0712$ resulting in $Q=13.8$. Hexagons are stable in the
regions enclosed by the solid lines and unstable outside. The dashed line represents 
the neutral curve. The results for H$_2$-Xe are obtained with $n_z=6$ and $n_q=5$  
\LB{fig:side-h2xe}.}
\end{figure}
\end{center}

The stability analysis of the H$_2$-Xe mixture also reveals an
oscillatory instability of the hexagon patterns at $\epsilon \sim 1$.
Within the $\epsilon$-range investigated, no such oscillatory
instability was found at the larger Prandtl numbers relevant for
CO$_2$ and SF$_6$. Unfortunately, it turns out that before the onset
of the oscillatory instability the hexagons already become unstable to
a side-band instability at half the hexagon wavelength, which will
always preempt the oscillatory instability. For the rolls we also find
an oscillatory instability. It is presumably related to the well-known
oscillatory instability of Boussinesq rolls \cite{CrWi89}. 

\begin{center}
\begin{figure}
\centering
\includegraphics[width=0.3\textwidth,angle=0]{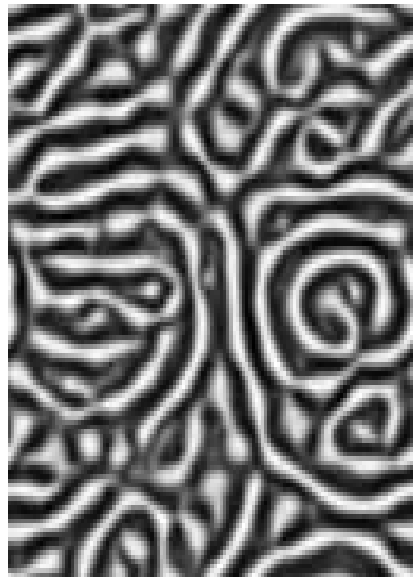} 
\caption{ 
Typical pattern for convection in H$_2$-Xe at $\epsilon=0.3$.
\LB{fig:pattern-h2xe}.  }
\end{figure}
\end{center}

For not too small $\epsilon$ generic initial conditions will not lead
to hexagonal patterns but rather to spiral defect chaos. A typical
snapshot of a pattern in this state is shown in
Fig.\ref{fig:pattern-h2xe} for $\epsilon=0.3$. Compared to the
patterns obtained in NB convection in SF$_6$ (cf.
Fig.\ref{fig:sf6-nb-ob}b) the patterns in H$_2$-Xe are less cellular
and do not show a large number of small bubbles. To quantify these and
other characteristics of the patterns we again apply the geometric
diagnostics introduced earlier \cite{RiMa06}.  

In Fig.\ref{fig:compact-h2xe} we show the normalized distribution
functions for the compactness of white and black components (cf.
Fig.\ref{fig:hist-compact-normalized}). Since there are only very few
black components their distribution function exhibits large
statistical fluctuations. Of particular interest is the distribution
function for the white components. It confirms the visual impression
that the number of compact components is significantly reduced
compared to the case of SF$_6$; in fact, while in SF$_6$ the maximum
of the distribution function is close to ${\mathcal C}=1$, in H$_2$-Xe
the absolute maximum is at ${\mathcal C}\sim 0.1$, which corresponds
to filamentary structures.

\begin{center}
\begin{figure}
\centering
\includegraphics[width=0.4\textwidth,angle=0]{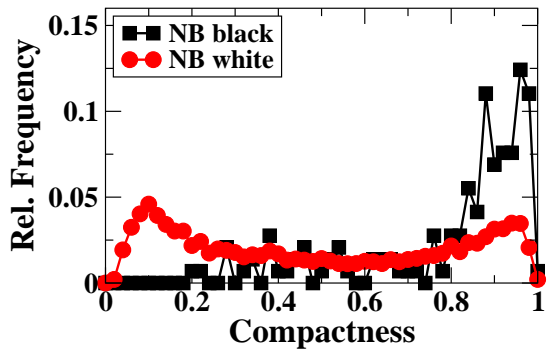}
\caption{ 
Distribution function of the compactness of closed contours 
for H$_2$-Xe ($Pr=0.17$, $\epsilon=0.3$)
\LB{fig:compact-h2xe}.  }
\end{figure}
\end{center}

The lack of small bubbles is demonstrated in more detail in the joint
distribution function for the contourlength and the compactness, which
is shown in Fig.\ref{fig:corr-cont-compt-h2xe}. Note that the view is
rotated compared to Fig.\ref{fig:corr-cont-comp}. The distribution
function is lacking the broad shoulder seen in NB convection in SF$_6$
(see Fig.\ref{fig:corr-cont-comp}b). Instead, the decay of the
distribution function towards long filamentary contours is quite
slow. 

\begin{center}
\begin{figure}
\centering
\includegraphics[width=0.4\textwidth,angle=0]{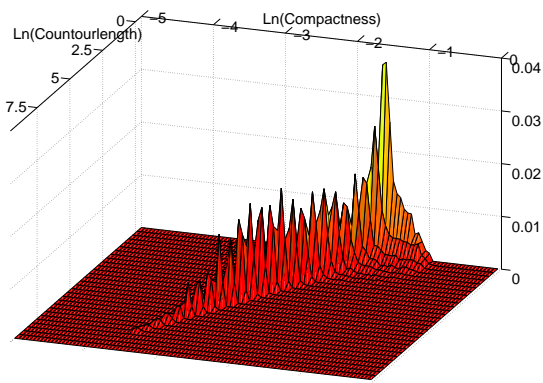}
\caption{ 
Joint distribution function for the contour length and 
the compactness of closed contours for H$_2$-Xe ($Pr=0.17$, $\epsilon=0.3$).
\LB{fig:corr-cont-compt-h2xe}.}
\end{figure}
\end{center}

To assess the spiral character of NB spiral defect chaos at these low
Prandtl numbers we show in Fig.\ref{fig:histo-winding} the
distribution function for the absolute value $|{\mathcal W}|$ of the
winding number for NB convection in H$_2$-Xe (at $\epsilon=0.3$ and
$PR=0.17$) as well as for Boussinesq and non-Boussinesq convection in
SF$_6$ (at $\epsilon=1.4$ and $Pr=0.8$). As had been noted
previously in the Boussinesq case \cite{EcHu97,RiMa06} the
distribution function is roughly consistent with exponential behavior.
In the NB case the exponential decays substantially faster than in the
Boussinesq case and spirals with winding numbers above ${|\mathcal
W}|=1$ are rare. In the Boussinesq case we had found that the decay
rate depends mostly on the Prandtl number, but only very little on
$\epsilon$ \cite{RiMa06}. Unfortunately, we do not have enough
non-Boussinesq data to investigate such trends in the $\epsilon$- and
$Pr$-dependence. However, it is worth noting that in the two
non-Boussinesq cases shown in Fig.\ref{fig:histo-winding} the decay
rates are essentially the same despite their substantial difference in
Prandtl number and both decays are much faster than that in the
Boussinesq case. Thus, possibly the impact of NB effects dominates
the dependence on the Prandtl number.

\begin{center}
\begin{figure}
\centering
\includegraphics[width=0.4\textwidth,angle=0]{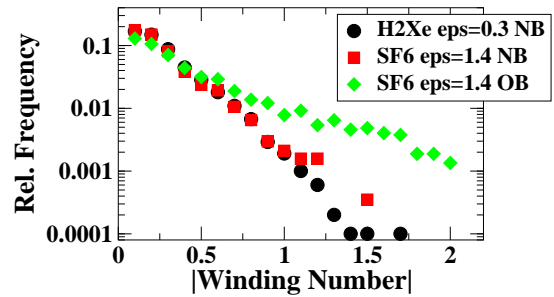}
\caption{ 
Distribution function for the absolute value of the winding number for
NB convection in H$_2$-Xe in comparison with convection in 
SF$_6$ in the NB and the OB case.
\LB{fig:histo-winding}.  }
\end{figure}
\end{center}

In NB convection in H$_2$-Xe, the distribution function for the number
of spirals is qualitatively similar to that in SF$_6$, which is shown
in Fig.\ref{fig:num-spiral} above. In particular, there are almost no
spirals with a winding number $|{\mathcal W}|>1$. Similarly, the
correlation between the arclength and the winding number of spirals
reveals that in NB convection in H$_2$-Xe there is no significant
trend towards Archimedean spirals. 

\section{Conclusion}
\LB{sec:conclusions}

In this paper we have studied non-Boussinesq convection in gases
(CO$_2$, SF$_6$, H$_2$-Xe) with Prandtl numbers ranging from $Pr\sim
1$ down to $Pr=0.17$ in experimentally relevant parameter regimes. We
have complemented a Galerkin stability analysis of hexagon patterns
with direct numerical simulations of the fluid equations to study
transitions between different hexagon patterns and to quantify the
impact of non-Boussinesq effects on spiral defect chaos.

We find that the reentrance of hexagons that we have identified
previously in non-Boussinesq convection in water \cite{MaRi05} also
occurs at low Prandtl numbers. As was the case at large Prandtl
numbers, compressibility is not neccessary for reentrance. Since, in
addition, the range of wavenumbers for which the reentrant hexagons
are stable differs significantly from that of the reentrant hexagons
observed in experiments on convection in SF$_6$ near the thermodynamic
critical point \cite{RoSt02}, the mechanisms underlying the two types
of restabilization of the hexagons is most likely different.
Reflecting the fact that in gases the non-Boussinesq effects increase
with increasing temperature the reentrance is shifted to lower values
of the Rayleigh number when the mean temperature is increased, opposite to
the behavior in water \cite{MaRi05}. In convection in water the
reentrant hexagons are stable only for wavenumbers below the critical
wavenumber. This trend becomes more pronounced with decreasing Prandtl
number. In fact, for the gas mixture with $Pr=0.17$ the wavenumber at
the stability limit of the hexagons decreases so rapidly with
increasing Rayleigh number that the range in Rayleigh number over
which the hexagons are stable becomes quite small. 

The comparison of our stability results with experiments on the
transition between hexagons and rolls in CO$_2$ \cite{BoBr91} shows
that this transition is not due to an amplitude or a side-band
instability. As a matter of fact, for the parameters of the
experimental system the hexagons do not undergo any linear amplitude
instability to rolls, contrary to the prediction of the weakly
nonlinear theory \cite{BoBr91}. We have performed detailed numerical
simulations with various lateral boundary conditions and confirm that
the transition is the result of the heterogeneous nucleation of rolls
at the side walls of the container, which then invade the whole system
if the Rayleigh number is sufficiently high. Our simulations suggest
that hexagons could be stabilized  well beyond the experimentally
observed transition point if the influence of the lateral walls can be
reduced by applying a spatially patterned forcing that drives hexagons
at the wall. Such a forcing can be achieved by localized heating
\cite{SeSc02} or by geometric patterning of the bottom plate
\cite{Bopriv}. Of course, the wavenumber of the forced hexagons would
have to be adjusted to lie in the stable range.

We have also investigated the stability of hexagons in H$_2$-Xe
mixtures with very small Prandlt number ($Pr=0.17$). There also stable
reentrant hexagons are possible, but they are restricted to a small
range in wavenumber ($q<q_c$) and Rayleigh number. Since for such
small Prandtl numbers Boussinesq hexagons are always side-band
unstable \cite{ClBu96}, the stability of the hexagons is a result of
the non-Boussinesq effects. 

A fascinating state that is characteristic for convection at low
Prandtl numbers is spiral defect chaos \cite{MoBo93}. We have studied
the influence of NB effects on spiral defect chaos in a set-up
corresponding to convection in SF$_6$ with a Prandtl number of $Pr=0.8$
and in H$_2$-Xe with a Prandtl number of $Pr=0.17$. To quantify the
differences between Boussinesq and non-Boussinesq spiral defect chaos
we have employed a recently suggested set of geometric diagnostics of
the patterns \cite{RiMa06}. 

As expected, in SF$_6$ and in H$_2$-Xe the non-Boussinesq effects
break the equivalence of up- and down-flows and, consequently,  the
mean number of pattern components (closed contours) corresponding to
down-flow differs significantly from that of the up-flow components.
More interesting is our finding that in SF$_6$ the total number of
components (Betti number \cite{GaMi04}) is more than twice as large in
the non-Boussinesq case than in the Boussinesq case for otherwise
equal parameters. We attribute this enhancement of cellular rather
than roll-like structures to the resonant triad interaction that is
introduced by the non-Boussinesq effects. 

Another striking difference between Boussinesq and non-Boussinesq
spiral defect chaos in SF$_6$ becomes apparent in the joint
distribution function for the contourlength of the components and
their compactness. While the Boussinesq case exhibits a strong
signature of target-like structures, the non-Boussinesq case displays
instead a marked increase in the number of small compact components
(`bubbles'). This trend towards a more cellular structure is also
apparent in the analysis of spirals. The number of spirals with a
winding number above 1/2 is much smaller in the non-Boussinesq than in
the Boussinesq case. In the correlations between the winding number
and the arclength of the components this is reflected by the missing
of large Archimedean spirals in the non-Boussinesq case. 

Interestingly, we do not find many small bubbles in our simulations of
H$_2$-Xe at $Pr=0.17$. In fact, the maximum of the distribution
function for the compactness of pattern components is shifted away
from ${\mathcal C}\lesssim 1$ to ${\mathcal C}\sim 0.1$, i.e. to
filamentary structures. We do not know whether this implies that the
tendency towards small bubbles is maximal at moderately small Prandtl
numbers, i.e. for $Pr \sim 1$, or whether it is due to the fact that for
computational reasons the simulations in H$_2$-Xe were performed at a
lower Rayleigh number ($\epsilon=0.3$) than in SF$_6$ ($\epsilon=1.4$). 

Due to computational limitations we have not been able to investigate
the transition from hexagons to spiral defect chaos at very small
Prandtl numbers. Since for very small Prandtl numbers there are
essentially no large spirals, the transition from hexagons to spiral
defect chaos may follow a different path than that observed
experimentally in CO$_2$, where large spirals were reported to arise
in the transition from hexagons to rolls \cite{BoBr91}. We surmise
that the geometric diagnostics that we have employed to characterize
the well developed spiral defect chaos would provide also valuable
insight into this transition. Since the diagnostics require a
substantial amount of data the transition would be best investigated
experimentally.

We deeply appreciate the support by W. Pesch and his students, who
have developed the codes we have used in this study \cite{DePe94a}. We
thank G. Ahlers for providing us with the code to determine the NB
coefficient. We have benefitted from discussions with G. Ahlers, G.
Gunaratne, K. Krishan, K. Mischaikow, W. Pesch, and M. Schatz. Support
from the Department of Energy (DE-FG02-92ER14303) and NSF
(DMS-9804673) is gratefully acknowledged.


\end{document}